\documentclass[%
 reprint,
 amsmath,amssymb,
 aps,
]{revtex4-2}
\usepackage{graphicx}
\usepackage{dcolumn}
\usepackage{bm}
\usepackage{hyperref}

\usepackage[caption=false]{subfig}

\usepackage{siunitx} 
\newcommand*{\suppl}{Supporting Information}

\begin{document}

\title{Enhancement of third harmonic generation induced by surface lattice resonances in plasmonic metasurfaces}
\author{Jeetendra Gour}
\affiliation{Friedrich Schiller University Jena, Institute of Applied Physics, Abbe Center of Photonics, Albert-Einstein-Str. 15, 07745 Jena, Germany}

\author{Sebastian Beer}
\affiliation{Friedrich Schiller University Jena, Institute of Applied Physics, Abbe Center of Photonics, Albert-Einstein-Str. 15, 07745 Jena, Germany}

\author{Alessandro Alberucci}
\affiliation{Friedrich Schiller University Jena, Institute of Applied Physics, Abbe Center of Photonics, Albert-Einstein-Str. 15, 07745 Jena, Germany}

\author{Uwe Detlef Zeitner}
\author{Stefan Nolte}
\affiliation{Friedrich Schiller University Jena, Institute of Applied Physics, Abbe Center of Photonics, Albert-Einstein-Str. 15, 07745 Jena, Germany}
\affiliation{Fraunhofer Institute for Applied Optics and Precision Engineering IOF, Albert-Einstein-Str. 7, 07745 Jena, Germany}

\begin{abstract}
We investigate experimentally Third Harmonic Generation (THG) from plasmonic metasurfaces consisting of two-dimensional rectangular lattices of centrosymmetric gold nano-bars. By varying the incidence angle and the lattice period, we show how Surface Lattice Resonances (SLRs) at the involved wavelengths are the major contributors in determining the magnitude of the nonlinear effects. A further boost on THG is observed when we excite together more than one SLR, either at the same or at different frequencies. When such multiple resonances take place, interesting phenomena are observed, such as maximum THG enhancement for counter-propagating surface waves along the metasurface, and cascading effect emulating a third-order nonlinearity.
\end{abstract}

\keywords{Nonlinear optics, third harmonic, plasmonics, surface lattice resonance, double resonance}

\maketitle
\section{Introduction}

The optical response of metallic nanostructures has been widely investigated due to its potential applications in several fields,  such as molecular sensing \cite{Mesch:2016}, frequency conversion \cite{Kauranen:2012,Lassiter:2014,Celebrano:2019}, and optical switching \cite{Min:2008}. For example, third-order nonlinear effects, such as Four Wave Mixing (FWM) and Third Harmonic Generation (THG), have been studied using pairs of coupled identical \cite{Hanke:2012} and asymmetric \cite{Harutyunyan:2012} nano-resonators, bow-tie nano-antennae \cite{Hentschel:2012}, U-shaped optical antennae \cite{Brien:2015}, and
nano-cavities encompassing metasurfaces \cite{Li:2022}. The main feature of plasmonics is the sub-wavelength confinement and the associated field enhancement \cite{Lassiter:2014}. The collective excitation of the free electrons at the interface of the metal (surface plasmon) is strongly coupled with the light, thus forming hybrid waves featuring a dispersion strongly differing from bulk dielectrics.
Such hybrid waves are called Surface Plasmon Polaritons (SPPs). There are different SPPs, spanning from Localized Surface Plasmons (LSPs) in isolated nanoparticles, to collective oscillations in a periodic arrangement of nanostructures \cite{Coutaz:1985}.
Such collective resonances are dubbed Surface Lattice Resonances (SLRs), and support extremely narrow resonances \cite{Kravets:2018}. Initially discussed in 1D metallic gratings \cite{Coutaz:1985}, SLRs are also relevant in 2D gratings \cite{Kravets:2018}, today referred to as plasmonic metasurfaces. Beyond boosting the linear response, SLRs enhance the nonlinear response of metallic periodic structures, both in 1D \cite{Coutaz:1985,Andreev:2006,Renger:2010} and in 2D \cite{Kauranen:2012} geometries. 
Second Harmonic Generation (SHG) in the presence of SLRs has been thoroughly investigated in a plasmonic metasurface \cite{Czaplicki:2016,Michaeli:2017,Hooper:2018}, including the influence of multiple resonances \cite{Michaeli:2017,Stolt:2022,Beer:2022}. 
Third order effects focusing on the SLR have been previously studied in metallic films engraved with sub-wavelength holes \cite{Nezami:2015} or linear grooves \cite{Andreev:2006,Renger:2010,Genevet:2010}.
Equivalent reports stressing out the dependence of THG on the lattice periodicity, including the role of multiple resonances, are currently missing.\\

In this Letter, we show the influence of SLR on the Third Harmonic (TH) emitted in transmission mode using gold nano-bars placed on a transparent substrate (fused silica) and arranged in rectangular arrays of various spacings (i.e., different periods) along one direction. We also show how the simultaneous activation of multiple resonances further enhances the nonlinear effects. When compared to the SHG, in THG the four photon interaction implies a wider set of efficient multiple resonances owing to the different phase matching condition to be fulfilled by the surface waves \cite{ShamsMousavi:2019}.  

\section{Results and discussion}
\subsection{Sample fabrication and linear characterization}
\begin{figure}[h!]
\centering\includegraphics[width=0.45\textwidth]{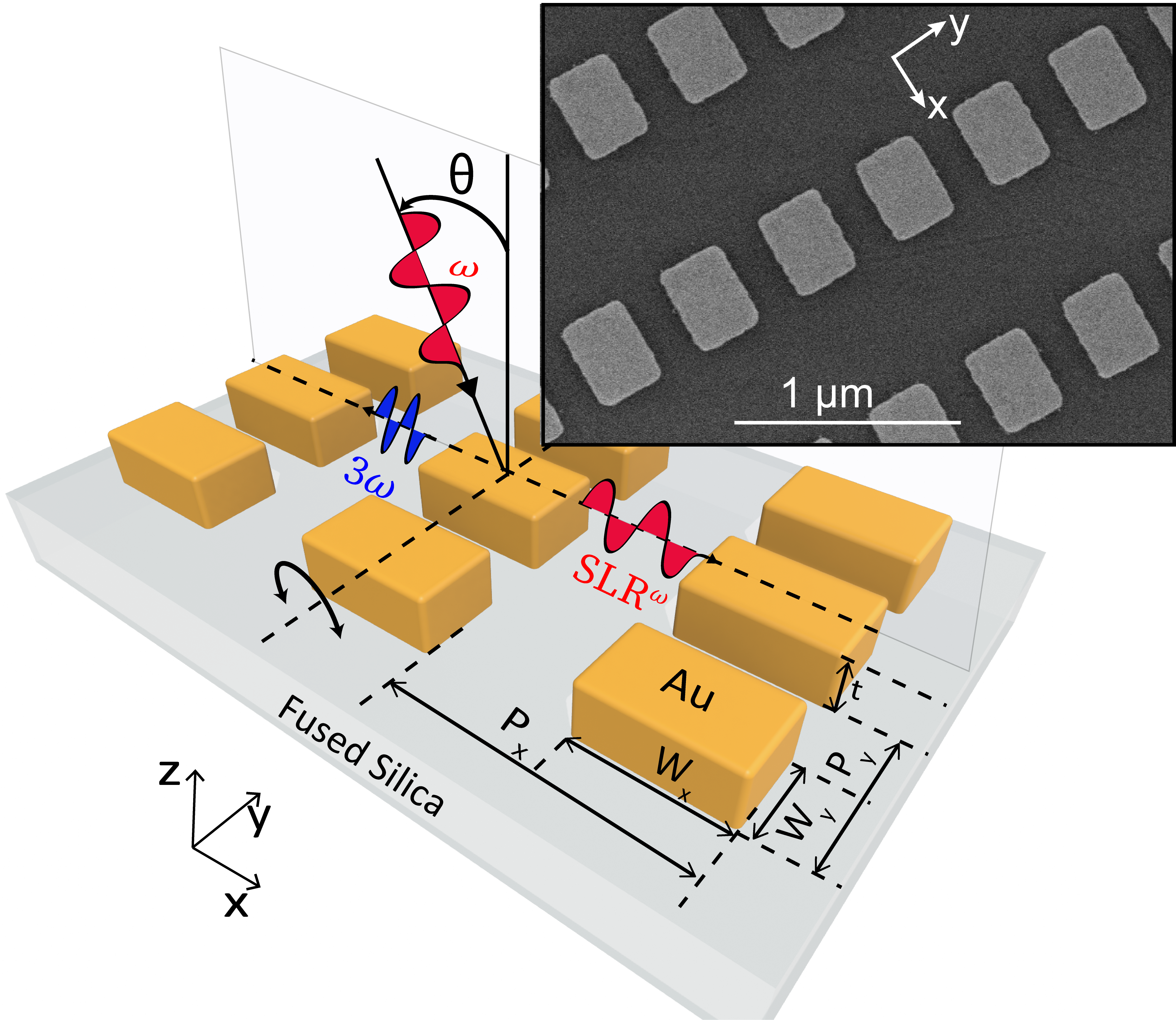}
\caption{Sample and excitation geometry. Inset: scanning electron microscope (SEM) image of the metasurface featuring $P_x =$ \SI{1000}{\nano \meter}. }
\label{Fig:Sample_design}
\end{figure}

Our samples are \SI{3}{\milli \meter}~$\times$~\SI{3}{\milli \meter} large metallic metasurfaces fabricated on a \SI{1}{\milli \meter} thick fused silica substrate, see Fig.~\ref{Fig:Sample_design}.
The nano-bars of dimensions $w_x=\SI{400}{\nano \meter}$, $w_y=\SI{300}{\nano \meter}$, and $t=\SI{50}{\nano \meter}$ are common to all the metasurfaces. 
The period $P_y$ in the $y$-direction ($P_y = $ \SI{500}{\nano \meter}) is fixed, whereas the period $P_x$ along the $x$-direction varies in the range $P_x \in$~[\SI{540}{\nano \meter}, \SI{1200}{\nano \meter}] with steps of \SI{20}{\nano \meter}. All the metasurfaces (i.e., for varying $P_x$) are fabricated on a single fused silica wafer of \SI{100}{\milli \meter} diameter. 
The high quality of the structure, mainly determined by the low roughness of the metallic surfaces, is confirmed by SEM imaging (see inset in Fig.~\ref{Fig:Sample_design}) and the very good agreement between measurements and simulations for the linear response (see \suppl) \cite{Beer:2022}.

The SLRs stem from the hybridization between localized plasmon and diffractive surface waves \cite{Kravets:2018}, the latter occurring close to the Rayleigh Anomalies (RAs).
In our case, the SLRs almost coincide with the RAs.
For a fixed period, RAs can be typically observed (thus, the surface wave excited) by varying the incidence angle $\theta$ \cite{Kravets:2018}, see Fig.~\ref{Fig:Sample_design}. As also sketched in Fig.~\ref{Fig:Sample_design}, the pump is impinging on the metasurface from the air side: indeed, in this configuration the measured damage threshold is 
higher.  Due to the asymmetric refractive index (air and glass),  we have two families of RAs, appearing when $\theta$ satisfies $\theta_{air} = {\arcsin}{\left(\pm 1 - \frac{m \lambda}{ P_x}\right)}$ and $\theta_{fs} = {\arcsin}{\left(\pm n_{fs} - \frac{m \lambda}{ P_x}\right)}$ respectively, where $m$ is the diffraction order, $\lambda$ is the vacuum wavelength, and $n_{fs}$ is the refractive index of fused silica.  
Given we are interested in SLRs occurring at any of the wavelength involved in the harmonic generation, in the following $\lambda$ can be either the Fundamental Wavelength (FW), the Second Harmonic (SH), or the Third Harmonic (TH).  

\subsection{Third harmonic generation}
\begin{figure}[ht!]
\centering
\includegraphics[width=0.5\textwidth]{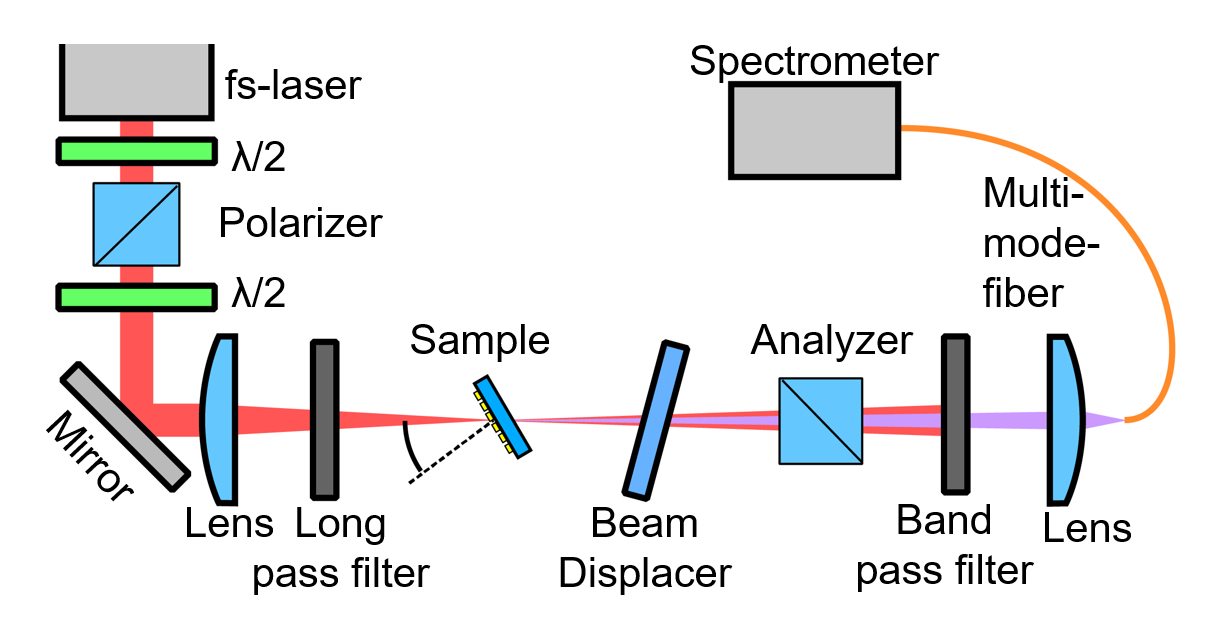}
\caption{Nonlinear characterization setup.}
\label{Fig:THG_setup}
\end{figure}

The experimental setup used for the nonlinear characterization  is shown in Fig.~\ref{Fig:THG_setup}. An ultrafast laser (pulse duration \SI{200}{\femto \second}, $\lambda =$ \SI{1032}{\nano \meter}, repetition rate  \SI{200}{\kilo \hertz}, average power \SI{125}{\milli \watt}) is pumping the metasurface, thus acting as the FW. The input power and linear polarization direction are adjusted via two Half Wave Plates (HWPs) and a polarizer. If not stated otherwise, the input polarization is parallel to the $x$ axis (TM polarization).
A plano-convex lens with a focal length of \SI{400}{\milli\meter} focuses the light to a beam diameter of approximately \SI{250}{\micro\meter} on the metasurface, leading to a peak fluency of \SI{2.55}{\milli\joule\per\centi\square\meter}, the latter corresponding to a peak intensity of \SI{12}{\giga \watt \per \square \centi \meter}.
The Rayleigh distance of the pump is much longer than the thickness of the substrate.
 
A long pass filter (FELH0900 by Thorlabs) is inserted in front of the sample to suppress the spurious generated light by the optical components preceding the metasurface. After the sample, two band-pass colored glass filters (2$\times$ FGUV11, \SI{275}{\nano \meter} - \SI{375}{\nano \meter}) attenuate the residual pump.
The TH signal is detected by a fiber-based spectrometer (\SI{193}{\nano \meter}-\SI{816}{\nano \meter}, Ocean HDX by Ocean Insight) with a back-thinned CCD.
We capture the TH emission in the $0^{th}$ order transmission, 
while changing the angle of incidence from \SI{-65}{\degree} to \SI{65}{\degree}, by rotating the metasurface along the $y$-axis. 
A rotatable plate of fused silica compensates for the beam transverse shift stemming from the rotation of the sample, thus maintaining the coupling to the multimode fiber constant when $\theta$ is varied.

\begin{figure}[ht!]
\centering
\includegraphics[width=0.5\textwidth]{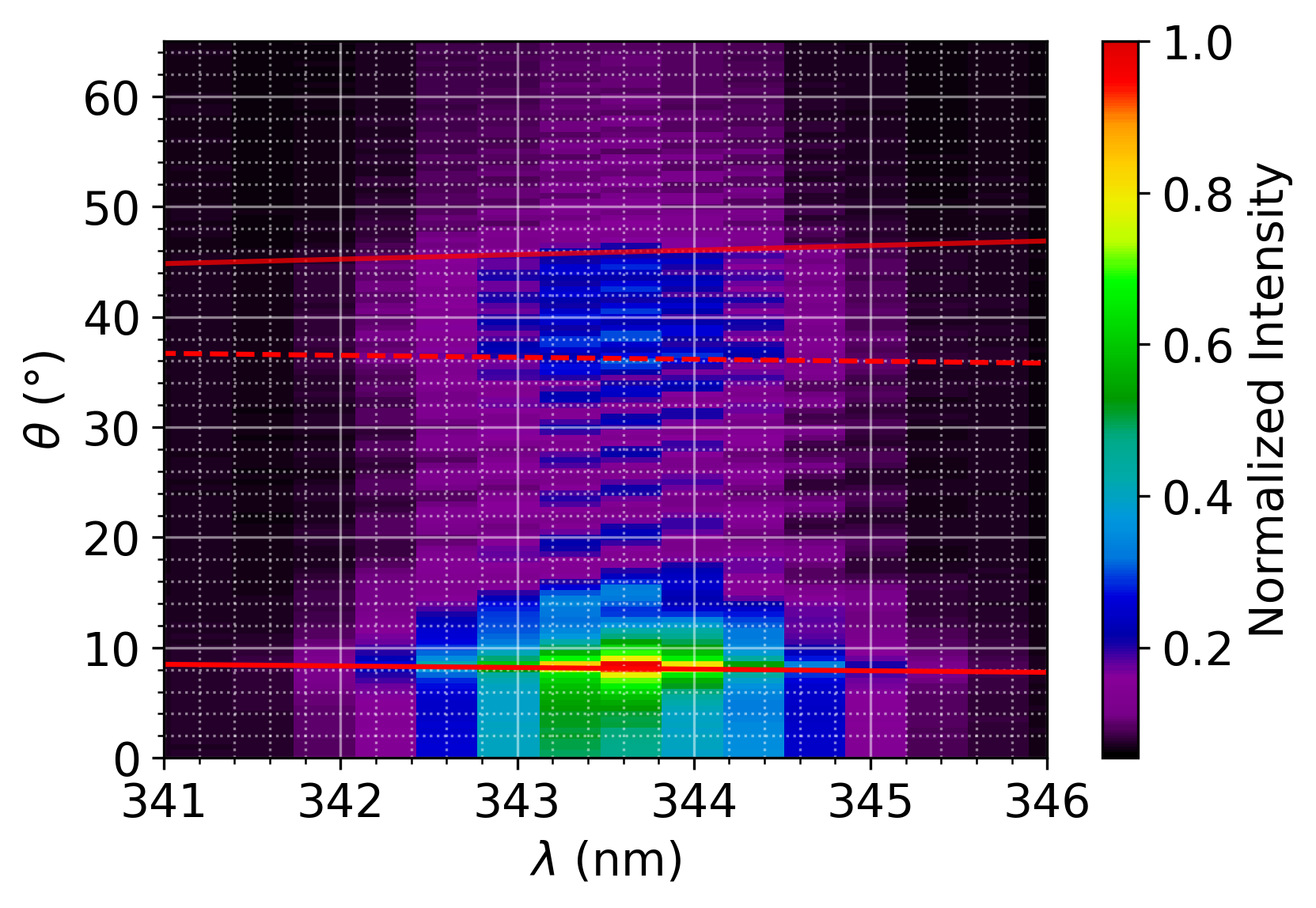}
\caption{Normalized acquired spectrum around the TH versus wavelength and incidence angle $\theta$ for $P_x=\SI{1200}{\nano\meter}$; solid and dashed red lines are air and glass RAs at FW, respectively. } 
\label{Fig:THG_spectrum_RA}
\end{figure}

A typical THG spectrum versus the incidence angle is shown in Fig.~\ref{Fig:THG_spectrum_RA} for a fixed pump spectrum of bandwidth \SI{10.5}{\nano\meter}. 
The signal is well localized (spectral bandwidth of about \SI{2}{\nano\meter}) around the TH ($\lambda=\SI{344}{\nano\meter}$) of the pump. 
The spectrum is symmetric within the experimental accuracy for positive and negative angles of incidence (see \suppl).
Due to the interference between TH emitted from the metasurface and from the glass slab, weak Maker fringes \cite{Maker:1962,Wang:1998, Gubler:2000} featuring a $\theta$-dependent visibility (e.g., a visibility of around \SI{26}{\percent} at $|\theta|=25^\circ$) are observed. The visibility is minimum when the signal from the metasurface is maximum, thus confirming the dominant role of the metasurface in the THG. For a given incidence angle $\theta$, the fringes can be eliminated by integrating the signal with respect to the wavelength (see \suppl). 
SLRs are fundamental in determining the shape of the emitted spectrum: three steep discontinuities in Fig. ~\ref{Fig:THG_spectrum_RA}, overlapping with the 1\textsuperscript{st} and 2\textsuperscript{nd} RAs in glass and in air, are observed in the emission. 

\begin{figure*}
\centering\includegraphics[width=1\textwidth]{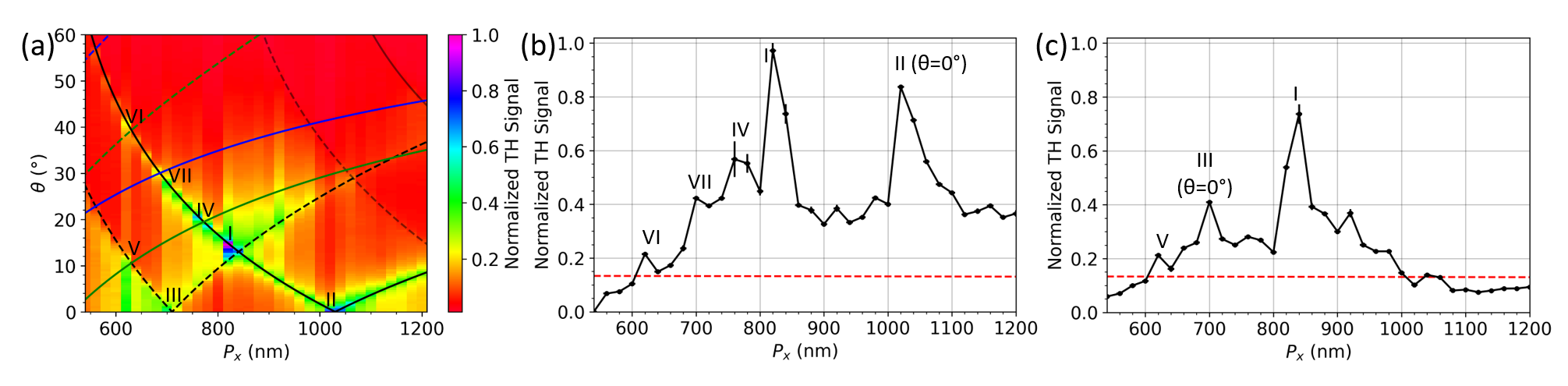}
\caption{(a) Emitted TH for TM input versus the period $P_x$ and the incidence angle $\theta$. 
Superposed solid and dashed lines are the RAs in air and in the glass, respectively; black, green and blue correspond the FW, SH and TH, respectively.   
Cross-section along (b) air-SLR [solid line in (a)] and (c) glass-SLR at FW [dashed line in (a)]. The red dashed horizontal line in (b-c) is the maximum TH signal from bare fused silica at $\theta=0^\circ$. The TH signal is normalized with respect to the global maximum. }
\label{Fig:THG_period_aoi}
\end{figure*}

An overview of the TH signal versus the incidence angle $\theta$ and versus the period $P_x$ (i.e., for different metasurfaces) is plotted in Fig.~\ref{Fig:THG_period_aoi}. 
The superposed lines correspond to the RAs, where the diffraction order propagating along the air (solid lines) or the glass interface (dashed lines). 
The surface waves related with the diffraction order +1 and -1 are degenerate only at normal incidence, splitting into two difference branches for skewed incidence. 
The branch bending towards longer periods corresponds to surface wave propagating along the same direction of the longitudinal component of the input wavevector. 
The wave propagates in the opposite direction for smaller periods. 
At the crossing points between two lines with opposite slopes, two counter-propagating surface waves \cite{Blau:1993} are thus simultaneously excited.
The importance of SLRs in THG is even more evident: the TH signal is maximized along the RAs, the effect being stronger at the RAs of the FW. 
Nonetheless, a SLR does not necessarily correspond to THG enhancement due to the role of the overlap integral, analogously to SHG \cite{Brien:2015,Beer:2022}.
Physically speaking, phase matching is the required condition to efficiently excite a surface plasmon at the fundamental wavelength (FW) \cite{ShamsMousavi:2019}. The nonlocal SPP at FW exploits the nonlinearity of the metasurface to excite other SPPs, but at higher harmonics. The presence of two families of SLRs due to the asymmetric environment increases the number of crossing points between RAs. Such points satisfy the multiple resonance condition \cite{Michaeli:2017}. We marked with Roman numbers up to $VII$ double resonance points where interesting physical effects are observed. 
The global maximum for the THG is situated in proximity of $I$, when two counter-propagating SLRs at the FW inside air and glass are simultaneously excited.
The conversion efficiency is in the order of $10^{-9}$ (see \suppl), similar to theoretical predictions for THG conversion efficiency from metallic nano-structures emitting in the UV-VIS range \cite{Scalora:15,Ciraci:15}. The
THG  is enhanced by a factor 8 with respect to the maximum emitted by the unstructured fused silica substrate, 
see the red dashed line in Fig.~\ref{Fig:THG_period_aoi} (more details can be found in the \suppl). 
The maximum THG is measured from the bare substrate at normal incidence ($\theta=\SI{0}{\degree}$) and at the focus position of the pump beam on the air-glass interface.

The THG is also significantly enhanced when two counter-propagating SLRs along the same interface are excited at normal incidence ($II$ at $P_x$=\SI{1020}{\nano\meter} with \SI{7}{\times}, $III$ at $P_x$=\SI{700}{\nano\meter} with \SI{4}{\times}).
Crossing points $IV$ (air SLR both at FW and SH), $V$ (glass SLR at FW, air SLR at SH) and $VI$ (air SLR at FW, glass SLR at SH), all show a relative maximum for the THG. This proves that a cascaded THG process, combining the Second Harmonic (SH) with the FW via a second-order nonlinearity, is non-negligible when the SH is close to resonance. Cascading effects in plasmonic metasurfaces have been theoretically discussed in the case of SLRs \cite{Doron:2019}, and experimentally proved using localized plasmons \cite{Celebrano:2019}. Finally, the resonance at the TH ($VII$) provides an appreciable but lower enhancement mainly due to the interband transition of gold. 
\begin{figure}[ht!]
\centering\includegraphics[width=0.5\textwidth]{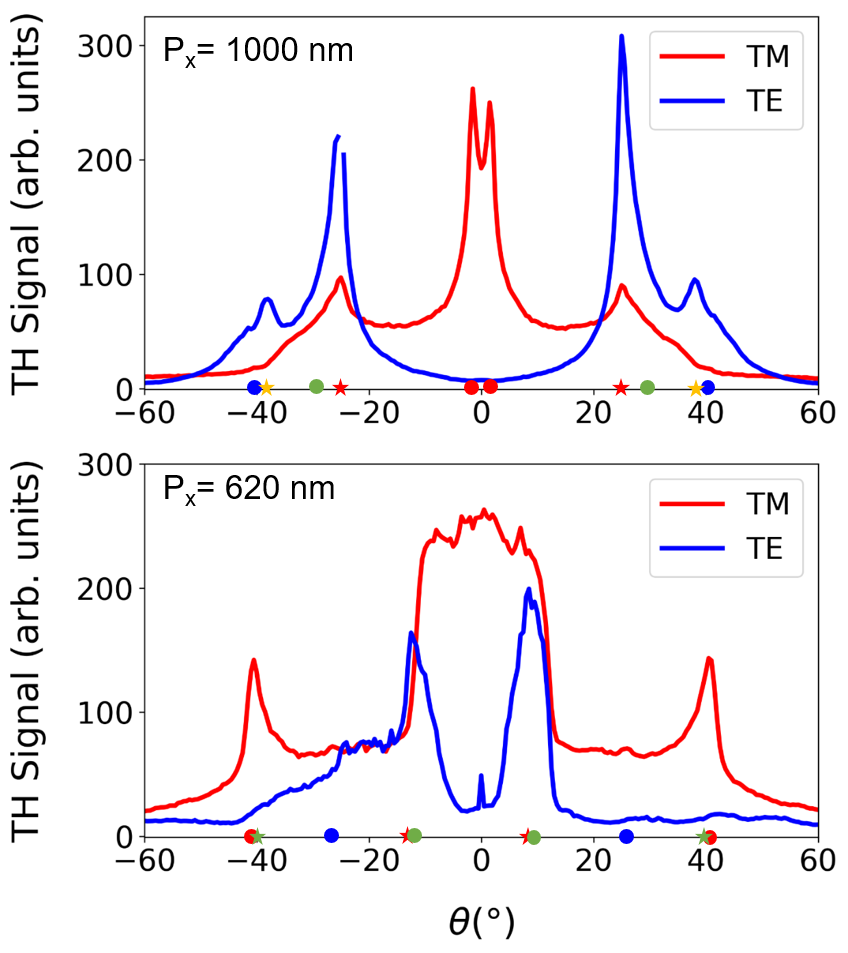}
\caption{TH signal versus the incidence angle $\theta$ for a period $P_x$ of 1000~nm (top panel) and 620~nm (bottom panel), measured for TE and TM linearly polarized inputs (see legend). Circle (star) markers on the axis correspond to air (glass) RAs. Red and gold color represent first ($|m|=1$) and second ($|m|=2$) order RAs at FW, respectively. Green and blue circles show first order RAs at the SH and the TH wavelengths, respectively.    }
\label{Fig:pol_THG_px_620_1000}
\end{figure}

We also investigated the dependence of THG on the polarization by rotating the input polarization and by placing an analyzer in front  of the spectrometer. Representative results for two different periods are shown in Fig.~\ref{Fig:pol_THG_px_620_1000}. 
The TH conserves the polarization of the fundamental, both for TM (Transverse Magnetic, polarized along $x$) and TE (Transverse Electric, polarized along $y$) linear inputs, with the cross-polarization terms being lower than \SI{2}{\percent}; such a value is determined by the accuracy ($\pm 2^\circ$) in the manual alignment of the HWP. Therefore, the cross-polarization results are not shown in Fig.~\ref{Fig:pol_THG_px_620_1000}. 
The THG response depends strongly on the polarization, since the linear response also changes \cite{Brien:2015,Beer:2022}. As demonstrated by Fig.~\ref{Fig:THG_period_aoi}(a), the most relevant contribution is given by the FW. When compared to TM polarization, for TE inputs the first order ($|m|=1$) SLR for the FW in air is very weak, whereas the corresponding SLR in glass is much stronger \cite{Auguie:2010,Beer:2022}.
For $P_x=1000~$nm, the TM signal is maximum near the normal incidence due to the presence of two resonant counter-propagating SLRs at the air interface (point II). For TE inputs, the maximum instead occurs around $\theta=25^\circ$, in correspondence with the RA at the glass interface of the FW. The second maximum occurs around the second order ($|m|=2$) RA of the FW at the glass interface and the first order ($|m|=1$) RA of the TH at the air interface. Noticeably, the effect of the resonance at the TH can be appreciated as a sudden change in the slope of the emitted signal.
For $P_x=$\SI{620}{\nano\meter}, TM signal shows a quasi-flat maximum around the normal incidence condition, the latter extending until the double resonance placed in point $V$. The presence of a tiny peak around the air RA of TH ($\theta=26.4^\circ$, corresponding to blue circle in the bottom panel of Fig.~\ref{Fig:pol_THG_px_620_1000}) and the crossing point $VII$ (in Fig.~\ref{Fig:THG_period_aoi}) suggest the non-negligible contribution of the TH SLR. In the case of TE, in $V$ the polarization shows  a sharp and distinct peak, whereas for $\theta=0^\circ$ a weak signal is measured. Finally, the THG spectrum retains the same shape as the input power is increased (see \suppl). The measured nonlinear signal increases with the third power of the input signal, with an exponent ranging from 2.83 to 2.98 (see \suppl).\\

\vspace*{1cm}
\section{Conclusions}
In summary, we have investigated experimentally the THG from periodic plasmonic metasurfaces  made of centrosymmetric gold nano-antenna and of different periodicity. We proved that the excitation of the SLR by the fundamental yields a steep increase in the THG, thus confirming the strict connection between linear and nonlinear properties in the case of perturbative nonlinear effects, in full analogy with the SHG case \cite{Brien:2015}. The maximum measured THG is almost an order of magnitude stronger than in bare fused silica, and it is achieved when two counter-propagating SLRs at the FW are simultaneously excited. An enhancement of TH potentially occurs in correspondence to the SLR of the SH, thus proving that cascading effects provide a tangible effect in generating the TH \cite{Huttunen:2018}. Our results hold valid for any material composing the substrate and for any shape of the resonator. Generalizations of the current work include the usage of more complex patterns \cite{Lim:2022}, or the engineering of the single nano-resonator \cite{Zhang:2016,Yang:2017} to improve the Q-factor of the metasurface. Another possible development is to repeat our measurements in a symmetric dielectric environment \cite{Kravets:2018,Bin:2021}: on one side, much larger Q-factors can be achieved; on the other side,  for a fixed angle of incidence two first-order counter-propagating surface waves cannot be excited simultaneously.

\begin{acknowledgments}
The authors thank the Deutsche Forschungsgemeinschaft for funding the project (CRC 1375 NOA, 398816777). 
We acknowledge the valuable support of Werner Rockstroh, Natali Sergeev, Detlef Schelle, Holger Schmidt and Daniel Voigt in the fabrication of the metasurfaces.
\end{acknowledgments}

\newpage


\bibliographystyle{unsrt}
\bibliography{references}

\clearpage
\pagebreak
\widetext
\begin{center}
\textbf{\large Supporting Information: \linebreak Enhancement of third harmonic generation induced by surface lattice resonances in plasmonic metasurfaces}
\end{center}
\setcounter{equation}{0}
\setcounter{figure}{0}
\setcounter{table}{0}
\setcounter{page}{1}
\setcounter{section}{0}
\makeatletter
\renewcommand{\theequation}{S\arabic{equation}}
\renewcommand{\thefigure}{S\arabic{figure}}
\renewcommand{\bibnumfmt}[1]{[S#1]}

In this supplemental document, we show scanning electron micrographs of a few  metasurfaces fabricated with high-quality and smooth edges, transmission spectrum of the metasurfaces, describe the post-processing and conversion efficiency of nonlinear TH signal, and provide the TH signal vs input power slope which confirms cubic nonlinearity of the third order process from a metasurface. We also show that the THG spectrum at different input-output polarization configurations maintains its shape when probed at different input powers.   

\section{Sample}
\begin{figure}[htbp]
\centering
\fbox{\includegraphics[width=0.9\linewidth]{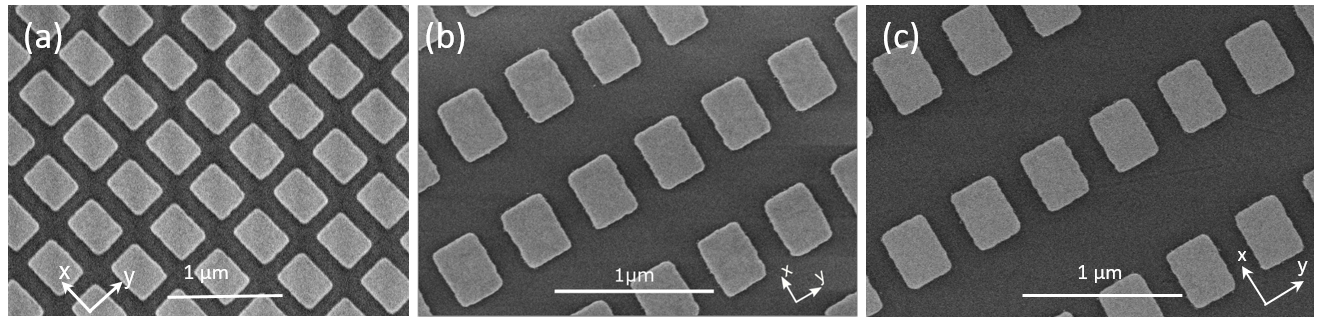}}
\caption{Scanning electron microscope images of the metasurfaces ((a) $P_x$ = \SI{540}{\nano \meter}, (b) $P_x$ = \SI{800}{\nano \meter}, and (c) $P_x$ = \SI{1000}{\nano \meter}) showing high-quality plasmonic nanostructures.}
\label{fig:sample_SEM}
\end{figure}

The fabricated metasurfaces on \SI{1}{\milli \meter} thick fused silica substrate show high quality plasmonic nanostructures which is evident from scanning electron microscope images of a few selected metasurfaces ($P_x$ = \SI{540}{\nano \meter}, \SI{800}{\nano \meter}, and \SI{1000}{\nano \meter}) in Fig.~\ref{fig:sample_SEM}. 

\section{Linear Optical Characterization}
\begin{figure}[htbp]
\centering
\fbox{\includegraphics[width=0.9\linewidth]{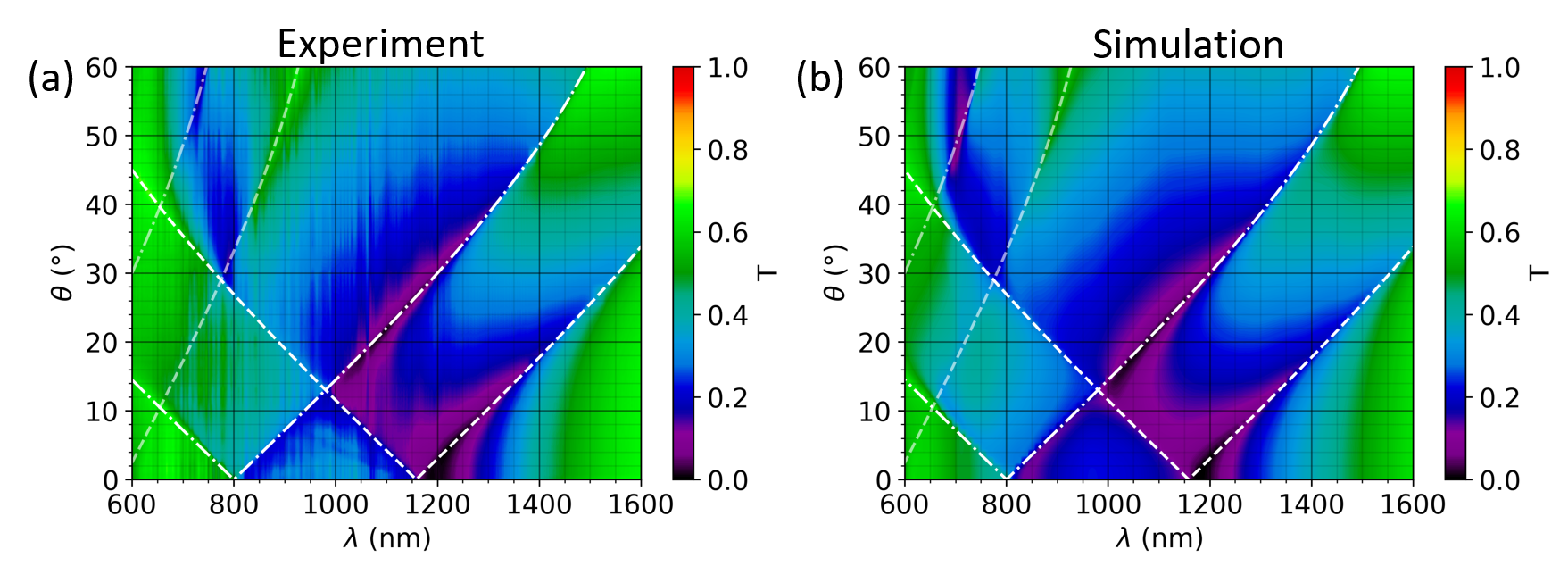}}
\caption{(a) Experimental and (b) simulated linear transmission spectrum of a metasurface at $P_x$ = \SI{800}{\nano \meter} with overlaid air/glass Rayleigh anomalies (dashed/dot-dashed white lines) of the first ($|m|=1$, bright lines) and second ($|m|=2$, faint lines) orders at TM input polarization.}
\label{fig:linear_response_sim_exp_px800}
\end{figure}

In Fig.~\ref{fig:linear_response_sim_exp_px800}, the simulated linear response of the metasurface shows quite good agreement with the measured transmission spectrum. It emphasizes the high quality of the fabricated metasurfaces. The linear transmission simulations are performed using rigorous coupled wave analysis (RCWA) by considering 225 harmonic waves. 

\begin{figure}[htbp]
\centering
\fbox{\includegraphics[width=0.9\linewidth]{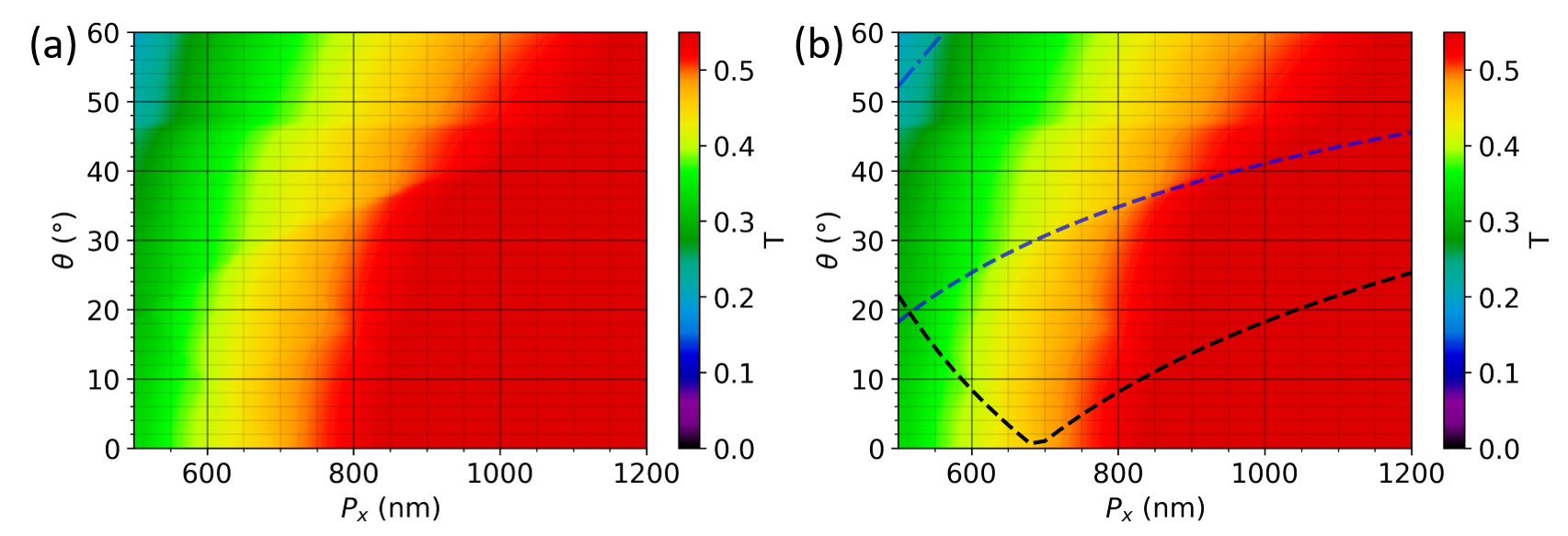}}
\caption{Simulated linear transmission spectrum of the metasurfaces at the third harmonic (\SI{344}{\nano \meter} wavelength for different metasurfaces (variation of $P_x$) with (a) and without (b) RA overlap at TM input polarization. Dashed blue/black lines correspond to first ($|m|=1$)/second($|m|=2$) order RAs.}
\label{fig:lnear_response_344_sim}
\end{figure}

\newpage
In Fig.~\ref{fig:lnear_response_344_sim} (a), the simulated response of the different metasurfaces shows a faint RA at the third harmonic wavelength while in Fig.~\ref{fig:lnear_response_344_sim} (b), it shows the perfect overlap of the calculated RAs.

\section{Nonlinear Optical Characterization}
\subsection{Post-processing}
\begin{figure}[htbp]
\centering
\fbox{\includegraphics[width=0.95\linewidth]{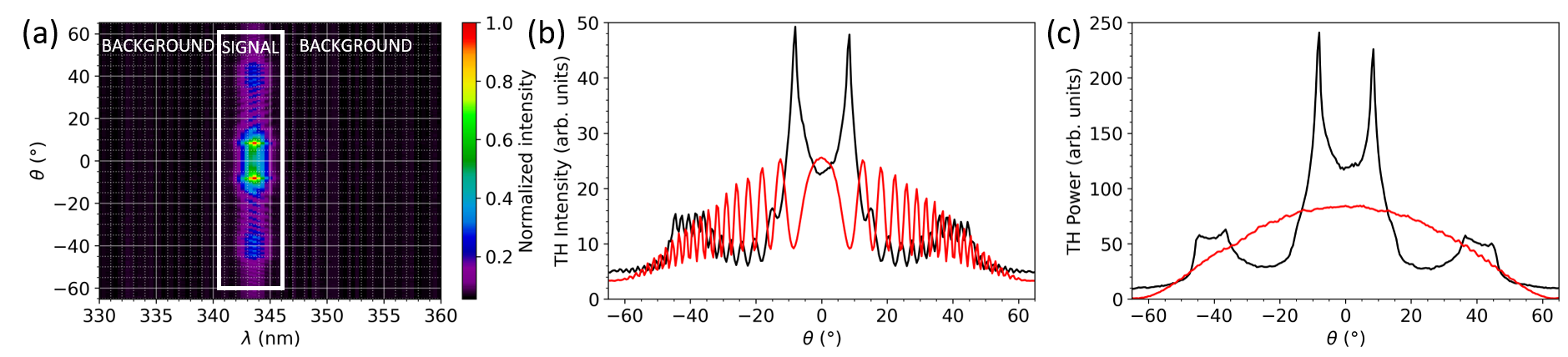}}
\caption{(a) Normalized THG spectrum of the metasurface ($P_x$ = \SI{1200}{\nano \meter}) on a fused silica substrate depending on the angle of incidence. Corresponding THG intensity at $\lambda$=\SI{343.6}{\nano\meter} (b) and the overall power (c)  measured with (black lines) and without (red lines) metasurface.}
\label{fig:post-processing}
\end{figure}

The finite thickness of the fused silica substrate leads to an oscillating pattern (Fig.~\ref{fig:post-processing}(a)) depending on the angle of incidence and wavelength, known as Maker fringes. These path length dependent fringes inside the substrate make the influence of SLRs on the THG hardly visible (Fig.~\ref{fig:post-processing}(b)). For a given angle $\theta$, the fringes can be removed by subtracting the average background noise level and integrating the TH spectrum (Fig.~\ref{fig:post-processing}(a) shown in white window). The power related TH signal shows no fringes Fig.~\ref{fig:post-processing}(c).
 
\newpage
\subsection{Conversion efficiency}
We estimated the conversion factor between the TH signal and the actual power. The absolute second harmonic power emitted from the metasurface was measured by comparing the signal of the spectrometer with the measured power from a calibrated detector. Such procedure was not possible for the TH due to the lower wavelengths. We estimated the absolute THG power under the assumption that the quantum efficiency of the spectrometer sensor is the main factor dictating the different detection efficiency at \SI{516}{\nano\meter} and at \SI{344}{\nano\meter}. This means that we are neglecting the wavelength dispersion in the detection system collecting the emitted light (absorption losses in the fiber, coupling into the fiber, diffraction losses at the grating). The global maximum TH signal in proximity of position I in figure 4 in the main text was 600 with a pump power of \SI{125}{\milli\watt}, which results in an estimated TH power of \SI{333}{\pico\watt} and a conversion efficiency of around $10^{-9}$.

\subsection{TH spectrum and power dependence}
\begin{figure}[htbp]
\centering
\fbox{\includegraphics[width=0.9\linewidth]{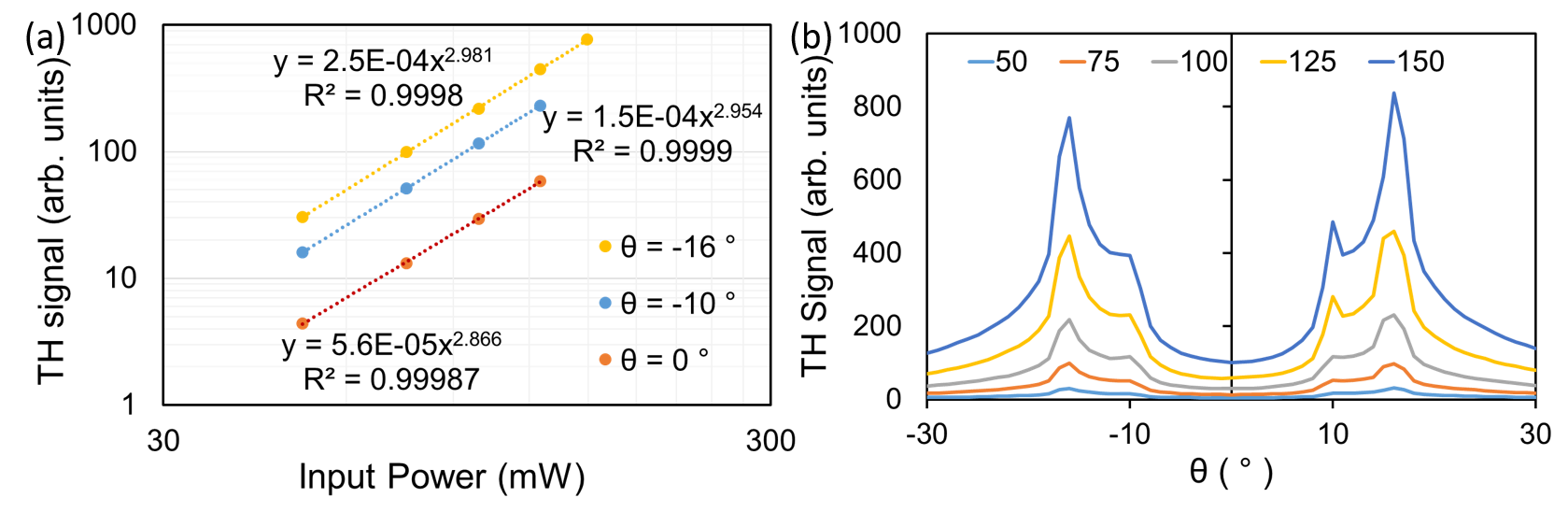}}
\caption{ (a) TH signal versus the input power for TM-TM configuration  for three incidence angles in a log-log scale. The dashed lines are power function fittings, the parameters being provided close to each curve. R-squared value shows how close the data points are to the fitted line.  (b) Angular dependence of the emission for different input powers.  
Here $P_x =$ \SI{800}{\nano \meter}.}
\label{fig:TH_signal_TM-TM}
\end{figure}

\begin{figure}[htbp]
\centering
\fbox{\includegraphics[width=0.9\linewidth]{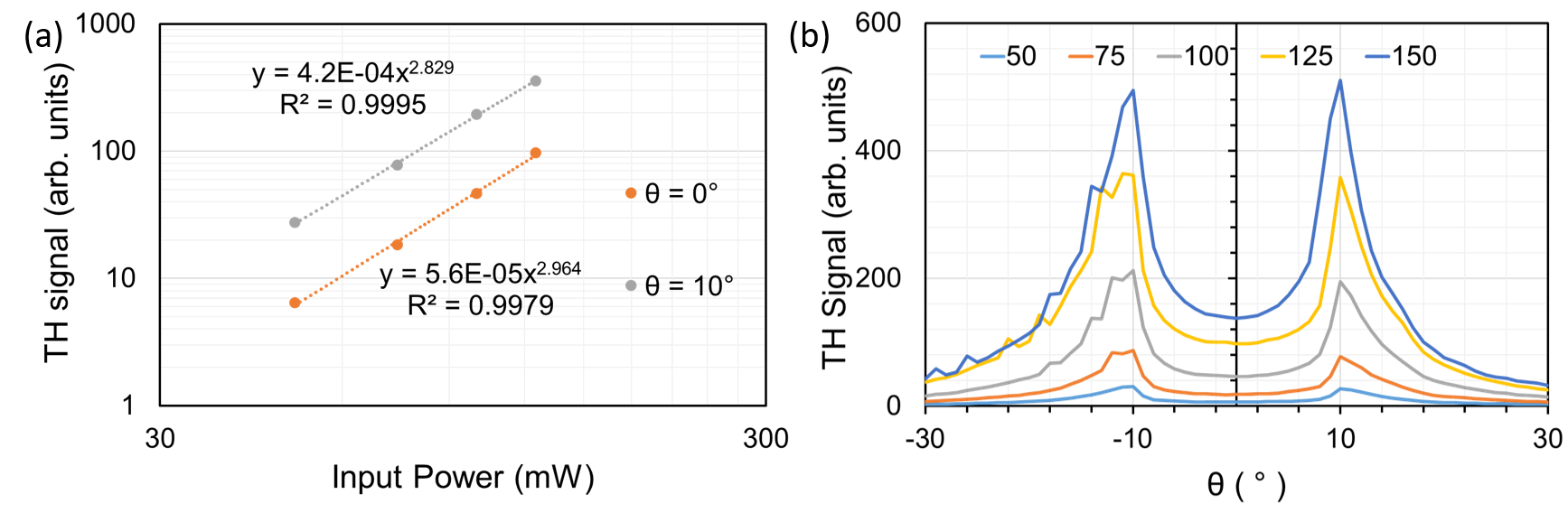}}
\caption{As in Fig.~\ref{fig:TH_signal_TM-TM}, but for TE-TE configuration.}
\label{fig:TH_signal_TE-TE}
\end{figure}

In Fig.~\ref{fig:TH_signal_TM-TM}, \ref{fig:TH_signal_TE-TE}, we show the output TH signal versus angle of incidence at different input powers which indeed follows the cubic relation and confirms the third order nonlinear process from the metasurface, the plots are shown in log scale. The standard error of all the slope fits are within the range of $0.05$.  The slight deviation of the slope value from 3 is not understood well, but suspect that it hints the presence of other nonlinear effects such as hot electron generation. We also show that for both TE-TE and TM-TM input-output polarization combinations, we
obtain enhanced THG at glass and air SLRs, respectively. 

\begin{figure}[ht!]
\centering
\fbox{\includegraphics[width=0.4\linewidth]{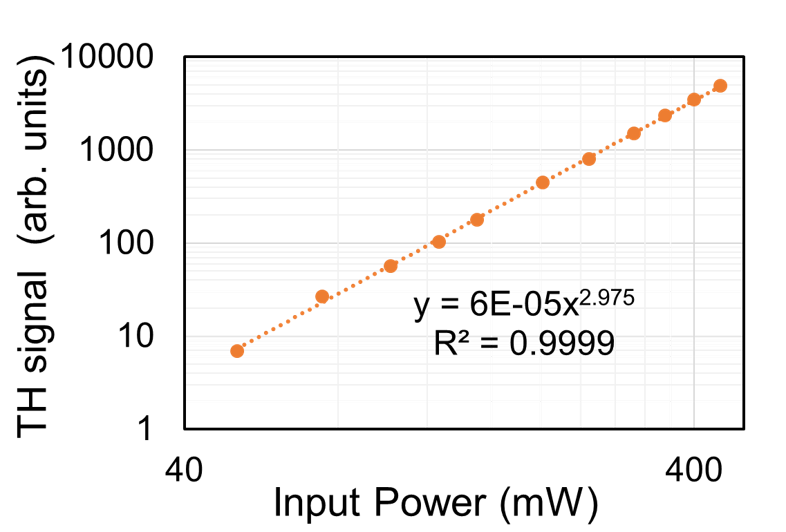}}
\caption{ TH signal versus input power slope for TE-TE configuration shows cubic nonlinearity in fused silica substrate at normal incidence.}
\label{fig:TH slope_fusi}
\end{figure}

In Fig.~\ref{fig:TH slope_fusi}, the output TH signal dependence on the input power of the ultrafast laser shows the slope of ~3 and cubic dependence on the input power for TE-TE configuration. 

\end{document}